\documentclass[%
reprint,
nofootinbib,
 amsmath,amssymb,
 aps,
]{revtex4-1}
\usepackage{stmaryrd}

\usepackage{graphicx}
\usepackage{dcolumn}
\usepackage{bm}
\usepackage{hyperref}

\usepackage[caption=false]{subfig}
\usepackage{graphicx}
\begin{document}

\preprint{APS/123-QED}

\title{Searching beyond the fiducial stochastic gravitational wave background in pulsar timing array data using likelihood reweighting}
\author{Heling Deng$^{1,2}$, Xavier Siemens$^{2}$, Rand Burnette$^{2}$ \\ \normalsize $^{1}$\textit{\small Department of Astronomy, Columbia University, New York, NY 10027, USA} \\ \normalsize $^{2}$\textit{\small Department of Physics, Oregon State University, Corvallis, OR 97331, USA} \\}

\begin{abstract}

{Since the recent announcements of evidence for a stochastic gravitational wave background from several pulsar timing array collaborations,} much effort has been devoted to explore features beyond 
{the fiducial Hellings-Downs background} including {those arising in} modified gravity theories and deterministic gravitational wave signals.
Inspired by previous studies, we propose a method to efficiently screen these models using likelihood reweighting based on {the fiducial model.} 
In order to alleviate the well-known unstable weight estimates in vanilla importance 
sampling, we implement reweighting for the second time making use of the kernel density estimation of the previously reweighted samples. We tested this method by analyzing three simulated datasets with an injected sinusoid signal applied to all pulsars. It is found that likelihood reweighting not only gives results compatible with those from full Bayesian analyses when the signal is subdominant, but is also able to recover the signal posterior to a reasonable accuracy in the presence of a rather strong signal. Given samples from the fiducial model, 
{this method} could bring an at least $\mathcal{O}(10)$-time speedup in analyzing new models. 

\end{abstract}
\maketitle

\section{Introduction}

In June 2023, the North American Nanohertz Observatory for Gravitational Waves (NANOGrav) reported compelling evidence for a nHz stochastic gravitational wave background (SGWB) in our Universe \cite{NANOGrav:2023gor}.  Meanwhile, the Chinese Pulsar Timing Array, the European Pulsar Timing Array (including data from the Indian Pulsar Timing Array), and the Parkes Pulsar Timing Array reported independent results at varying levels of significance \cite{Xu:2023wog,EPTA:2023fyk,Reardon:2023gzh}. The recent announcement by the MeerKAT Pulsar Timing Array further confirmed this discovery \cite{miles2025meerkat}. 

A promising source of the detected SGWB is the combined emission of gravitational radiation from an ensemble of inspiraling supermassive black hole (SMBH) binaries.  A SGWB in the nHz band could also be generated by physics in the early universe, such as inflation \cite{Grishchuk:1974ny,Starobinsky:1979ty,Rubakov:1982df,Fabbri:1983us,Vagnozzi:2023lwo}, phase transitions \cite{Kosowsky:1991ua,Kosowsky:1992rz,Kosowsky:1992vn,Kamionkowski:1993fg,Caprini:2007xq,Huber:2008hg,Hindmarsh:2013xza,Giblin:2013kea,Giblin:2014qia,Hindmarsh:2015qta,NANOGrav:2021flc} and cosmic strings \cite{Vilenkin:1981bx,Damour:2004kw,Buchmuller:2020lbh,Ellis:2020ena,Blanco-Pillado:2021ygr,Hindmarsh:2022awe}. See also Refs. \cite{NANOGrav:2023hvm,Ellis:2023oxs,Figueroa:2023zhu}. These cosmological models are also under active investigation.

Regardless of the constituent(s), the fiducial model of the SGWB describes a homogeneous and isotropic red noise process that not only has a spectrum common among all pulsars, but is also spatially correlated among pulsar pairs characterized by the Hellings-Downs (HD) correlations \cite{Hellings:1983fr}, which is a decisive signature in the pulsar timing array (PTA) predicted by general relativity. Modified theories of gravity can cause deviations from the expected correlation pattern \cite{NANOGrav:2023ygs,Chen:2023uiz,Bernardo:2023zna,Bi:2023ewq, Agazie:2024qnx}. Moreover, deterministic gravitational wave signals in addition to the stochastic background may be present in the PTA data. An example is continuous waves emitted by the brightest SMBH binary in the sky. The detection of an individual binary would provide direct evidence for the existence of SMBH binaries and shed light on the source of the stochastic background itself \cite{Sesana:2008xk,Rosado:2015epa,Kelley:2017vox,Becsy:2022pnr,NANOGrav:2023bts,IPTA:2023ero}. 

Models beyond the fiducial HD background typically introduce new parameters. While it is straightforward to assess the significance of new models by performing full Bayesian inference via, e.g., Markov chain Monte Carlo (MCMC), it is also possible to exploit the results of existing searches to efficiently explore the posteriors of additional parameters in beyond-HD models. 
This is possible when the new features do not dominate the HD background, i.e., when the new model is a ``perturbed version" of the fiducial HD model.

In this paper, we investigate models beyond the fiducial SGWB in the PTA data using likelihood reweighting, which is based on importance sampling.\footnote{{Likelihood reweighting is closely related to the stepping-stone sampling algorithm introduced in Ref. \cite{xie2011improving} (applications in searching for gravitational waves can be found in, e.g., Refs. \cite{Maturana-Russel:2018yos, Zahraoui:2024oqa}), which was later adapted for model selection in Ref. \cite{vitoratou2017thermodynamic}.}} This method was applied in the LIGO context to study higher-order gravitational wave modes \cite{Payne:2019wmy}, and the eccentricity of a black hole binary orbit \cite{Romero-Shaw:2019itr}.  {A brief introduction can also be found in Ref. \cite{Biscoveanu:2021cpd}}. In the PTA context, it was found in Ref. \cite{Hourihane:2022ner} that the method is a powerful tool for studying the HD model starting from a model with an uncorrelated background. Building on these studies, we used likelihood reweighting to explore beyond-SGWB features in simulated PTA datasets. {Using a global sinusoid as a toy model, we show that likelihood reweighting efficiently and accurately predicts the parameters of the sinusoid as well as the SGWB when the sinusoid is relatively weak, and that it gradually fails to predict the SGWB parameters as the signal amplitude increases. Even in the latter case, reweighting still provides accurate parameter recovery of the global signal.} In addition, we find that this technique provides at least a $\mathcal{O}(10)$-time speedup relative to a full Bayesian search.

{Furthermore, in order to estimate the SGWB parameters more accurately in the presence of a strong deterministic signal, we propose to perform reweighting for the second time based on the results of the first reweighting. Adopting this step helps to temper the well-known unstable weight estimates in importance sampling.}

The remainder of the paper is organized as follows. In Section \ref{section 2} we describe the general formulation of the method. In Section \ref{section 3} we briefly introduce the PTA likelihood, and then demonstrate how likelihood reweighing can be applied in searching for new signals. Conclusions are summarized and discussed in Section \ref{section 4}.  

\section{Methodology\label{section 2}}

The idea of likelihood reweighting is based on importance sampling, which is a simple modification of the Monte Carlo method for Bayesian inference. 
{Suppose we would like to use an observation dataset to constrain some physical models. Consider a ``proposal model" described by parameters $\boldsymbol{x}$, whose posterior is relatively easy to find by MCMC; and a competing ``target model" with the same parameters $\boldsymbol{x}$ and some additional parameters $\boldsymbol{y}$.}
If $\boldsymbol{y}$ have a much less impact on the dataset than $\boldsymbol{x}$, then the target effectively behaves like the proposal plus a small perturbation. Rather than directly analyzing the target, it would be more efficient if we can extract the information about $\boldsymbol{y}$ by exploiting what we (already) know about the proposal. This can be achieved by assigning proper weights to the proposal's posterior samples.

\subsection{Reweighting}

Let $\mathcal{L}$ ($\mathcal{L}_{0}$) be the likelihood of the target (proposal), and $Z$ ($Z_{0}$) be the corresponding evidence. The proposal's posterior is then given by $p_{0}(\boldsymbol{x})=\mathcal{L}_{0}(\boldsymbol{x})\pi_{0}(\boldsymbol{x})/Z_{0}$, where $\pi_{0}(\boldsymbol{x})$ is the prior.\footnote{Strictly speaking, the posterior and the likelihood should be written as $p_{0}(\boldsymbol{x}|\text{data})$ and $\mathcal{L}_{0}(\text{data}|\boldsymbol{x})$ respectively. For simplicity we ignore the ``data" part in this Section.} Assuming that the target's prior is $\pi_{0}(\boldsymbol{x})\pi(\boldsymbol{y})$, i.e., the priors for $\boldsymbol{x}$ in the two models are the same, its posterior can then be expressed as
\begin{equation}
p(\boldsymbol{x},\boldsymbol{y})\propto\mathcal{L}(\boldsymbol{x},\boldsymbol{y})\pi_{0}(\boldsymbol{x})\pi(\boldsymbol{y})\propto w(\boldsymbol{x},\boldsymbol{y})p_{0}(\boldsymbol{x})\pi(\boldsymbol{y}),\label{eq:pos_x_y}
\end{equation}
where we have defined the weight $w$ as the likelihood ratio:
\begin{equation}
w(\boldsymbol{x},\boldsymbol{y})\equiv\frac{\mathcal{L}(\boldsymbol{x},\boldsymbol{y})}{\mathcal{L}_{0}(\boldsymbol{x})}.
\end{equation}
Therefore, in order to obtain the target's posterior, we simply need to assign a weight 
\begin{equation}
w_i=w(\boldsymbol{x}^{(i)},\boldsymbol{y}^{(i)}),\label{eq:weight}
\end{equation}
to each MCMC sample of the proposal's posterior, where $\boldsymbol{x}^{(i)}$ denotes the $i$th sample, and $\boldsymbol{y}^{(i)}$ represents a sample randomly drawn from the prior $\pi(\boldsymbol{y})$.  Obtaining $p(\boldsymbol{x},\boldsymbol{y})$ this way can be more efficient than performing a full Bayesian search. For one thing, the target's likelihood can now be computed fewer times, because the proposal's samples have been thinned out before the reweighting. For another, the weights can be computed in parallel, in which case the wall time is significantly reduced. 

Moreover, this method can also help in model selection because the weights can give us the Bayes factor between the two models. To this end, the target's evidence can be written as
\begin{align}
Z & =\int\mathcal{L}(\boldsymbol{x},\boldsymbol{y})\pi_{0}(\boldsymbol{x})\pi(\boldsymbol{y})\text{d}\boldsymbol{x}\text{d}\boldsymbol{y}\\
 & =\int\mathcal{L}(\boldsymbol{x},\boldsymbol{y})\pi_{0}(\boldsymbol{x})\pi(\boldsymbol{y})\frac{p_{0}(\boldsymbol{x})}{\mathcal{L}_{0}(\boldsymbol{x})\pi_{0}(\boldsymbol{x})/Z_{0}}\text{d}\boldsymbol{x}\text{d}\boldsymbol{y}\\
 & =Z_{0}\int\frac{\mathcal{L}(\boldsymbol{x},\boldsymbol{y})}{\mathcal{L}_{0}(\boldsymbol{x})}\pi(\boldsymbol{y})p_{0}(\boldsymbol{x})\text{d}\boldsymbol{x}\text{d}\boldsymbol{y}\\
 & \approx\frac{Z_{0}}{N}\sum_{i=1}^{N}w_i,\label{eq:Z}
\end{align}
where $N$ is the sample number in the proposal's MCMC chain. Therefore, the evidence ratio of the two models, or the Bayes factor, is the averaged weight over $N$ samples:
\begin{equation}
\mathcal{B}\equiv\frac{Z}{Z_{0}}\approx\bar{w}.
\end{equation}
The standard error of the Bayes factor is\footnote{{Note the difference between this expression and the one in Ref. \cite{Hourihane:2022ner} [Eq. (13)], where $N$ was mistakenly written as $N_{\rm eff}$ defined below.}}
\begin{equation}
\Delta\mathcal{B}=\frac{\sigma_{w}}{\sqrt{N}},
\end{equation}
where $\sigma_{w}$ is the standard deviation of the weights.

The performance of this technique relies on the similarity of the two models. When the proposal is a poor approximation to the target, there is parameter space where proposal's samples are rare, whereas the target's posterior density there is non-negligible. This is well known to cause unstable weight estimates with large variance. Furthermore, $\pi(\boldsymbol{y})$ is usually assumed to be uniform since we do not know much about these additional parameters, so the search on $\boldsymbol{y}$ is vanilla Monte Carlo, which requires a relatively large number of samples. Hence, the dimension of $\boldsymbol{y}$ cannot be too large (in the experiments below, the dimension of $\boldsymbol{x}$ is 42, while the dimension of $\boldsymbol{y}$ is 3). The efficiency of the resampling can be characterized by $N_{\text{eff}}/N$, where $N_{\text{eff}}$ is the ``effective sample number'', given by
\begin{equation}
N_{\text{eff}}=\frac{\left(\sum_{i=1}^{N}w_{i}\right)^{2}}{\sum_{i=1}^{N}w_{i}^{2}}=\left[1+\left(\frac{\sigma_w}{\bar{w}}\right)^2\right]^{-1}N.
\end{equation}
$N_{\text{eff}}$ tells us whether the target's posterior is sufficiently sampled.\footnote{{In the case of low resampling efficiency, $N_{\rm eff}$ can be estimated as $(\Delta\mathcal{B}/\mathcal{B})^2$.}} A threshold of $N_{\text{eff}}$ is usually determined empirically. If the threshold is not reached, it is straightforward to generate more weighted posterior samples from the proposal's MCMC chain. When nontrivial new parameters ($\boldsymbol{y}$) are involved, a low efficiency is almost guaranteed because the weights are sensitive to the sample values of these parameters. We will deal with this problem in the next subsection.

\subsection{Second reweighting}

In order to enhance the performance of the reweighting, we have further adopted the idea of adaptive importance sampling (see, e.g., Ref. \cite{bugallo2017adaptive} for a review), which is to find a better proposal model (usually iteratively) based on the previously reweighted samples. After obtaining the posterior of $\boldsymbol{x}$ and $\boldsymbol{y}$ using Eq. (\ref{eq:pos_x_y}), which gives us $N$ reweighted samples, we can use kernel density estimation (KDE) to approximate the marginal posterior of $\boldsymbol{y}$:
\begin{equation}
f_{\text{KDE}}(\boldsymbol{y})\approx p(\boldsymbol{y}).
\end{equation}
Now consider a likelihood given by $f_{\text{KDE}}(\boldsymbol{y})$, based on which we easily draw $N$ samples from $p_{\text{KDE}}(\boldsymbol{y})=f_{\text{KDE}}(\boldsymbol{y})\pi(\boldsymbol{y})$ assuming a prior $\pi(\boldsymbol{y})$. We then construct a new proposal model with likelihood
\begin{equation}
\label{fKED}
\mathcal{L}_{1}(\boldsymbol{x},\boldsymbol{y})=\mathcal{L}_{0}(\boldsymbol{x})f_{\text{KDE}}(\boldsymbol{y}).
\end{equation}
Note that $\boldsymbol{x}$ and ${\boldsymbol{y}}$ are independent. This new proposal model should be better than the first one because it encodes some information about parameters ${\boldsymbol{y}}$.

Similar to Eq. (\ref{eq:pos_x_y}), the target's posterior can be found by reweighting the samples from $p_{0}(\boldsymbol{x})$ and $p_{\rm KDE}(\boldsymbol{y})$, because 
\begin{equation}
p(\boldsymbol{x},\boldsymbol{y})\propto\mathcal{L}(\boldsymbol{x},\boldsymbol{y})\pi_{0}(\boldsymbol{x})\pi(\boldsymbol{y})\propto w^\prime(\boldsymbol{x},\boldsymbol{y})p_{0}(\boldsymbol{x})p_{\rm KDE}(\boldsymbol{y}),\label{eq:pos_x_y2}
\end{equation}
where we have defined the weight $w^{\prime}$ as the likelihood ratio:
\begin{equation}
w^{\prime}(\boldsymbol{x},\boldsymbol{y})\equiv\frac{\mathcal{L}(\boldsymbol{x},\boldsymbol{y})}{\mathcal{L}_{1}(\boldsymbol{x},\boldsymbol{y})}.\label{eq:weight-1}
\end{equation}
To each sample of the new proposal's posterior $p_{0}(\boldsymbol{x})p_{\rm KDE}(\boldsymbol{y})$, we assign a weight
\begin{equation}
w^\prime_i=w^\prime(\boldsymbol{x}^{(i)},\boldsymbol{y}^{(i)}),
\end{equation}
where $\boldsymbol{x}^{(i)}$, as in Eq. (\ref{eq:weight}), denotes the $i$th MCMC sample of the first proposal's posterior $p_{0}(\boldsymbol{x})$, and $\boldsymbol{y}^{(i)}$ now represents a sample randomly drawn from $p_{\text{KDE}}(\boldsymbol{y})$. 

Let $Z_{1}$ be the evidence of this new proposal. Similar to Eq. (\ref{eq:Z}), the target's evidence can be written as
\begin{align}
Z & =\int\mathcal{L}(\boldsymbol{x},\boldsymbol{y})\pi_0(\boldsymbol{x})\pi(\boldsymbol{y})\text{d}\boldsymbol{x}\text{d}\boldsymbol{y}\\
 & =\int\mathcal{L}(\boldsymbol{x},\boldsymbol{y})\pi_0(\boldsymbol{x})\pi(\boldsymbol{y})\frac{p_{0}(\boldsymbol{x})p_{\text{KDE}}(\boldsymbol{y})}{\mathcal{L}_{1}(\boldsymbol{x},\boldsymbol{y})\pi_{0}(\boldsymbol{x})\pi(\boldsymbol{y})/Z_{1}}\text{d}\boldsymbol{x}\text{d}\boldsymbol{y} \label{Z1}\\
 & \approx\frac{Z_{1}}{N}\sum_{i=1}^{N}w^{\prime}_i\label{Z2}.
\end{align}
Rather than $Z/Z_1$, $Z/Z_0$ is what we need for model selection, because the new proposal model is usually not physically interesting; it is an auxiliary model that we construct to connect the first proposal and the target. Note that, by definition, $f_{\text{KDE}}(\boldsymbol{y})$ is a probability distribution; it is almost equal to $p_{\text{KDE}}(\boldsymbol{y})$ if the KDE has a sufficiently small bandwidth. Hence
\begin{align}
Z_{1} & =\int\mathcal{L}_{1}(\boldsymbol{x},\boldsymbol{y})\pi_{0}(\boldsymbol{x})\pi(\boldsymbol{y})\text{d}\boldsymbol{x}\text{d}\boldsymbol{y}\\
 & =\int\mathcal{L}_{0}(\boldsymbol{x})\pi_{0}(\boldsymbol{x})\text{d}\boldsymbol{x}\int f_{\text{KDE}}(\boldsymbol{y})\pi(\boldsymbol{y})\text{d}\boldsymbol{y}\\
 & \approx Z_{0}\int p_{\text{KDE}}(\boldsymbol{y})\pi(\boldsymbol{y})\text{d}\boldsymbol{y}.
\end{align}
The Bayes factor between the target and the first proposal then becomes 
\begin{equation}
\mathcal{B}=\frac{Z}{Z_{0}}\approx\rho \bar{w^{\prime}},
\end{equation}
where we have defined $\rho\equiv \left[\int p_{\text{KDE}}(\boldsymbol{y})\pi(\boldsymbol{y})\text{d}\boldsymbol{y}\right]^{-1}$. If $\pi(\boldsymbol{y})$ is uniform, $\rho$ is simply the prior density. 

For efficiency, we obtain $f_{\text{KDE}}(\boldsymbol{y})$ from one dimensional marginal posteriors of parameters $\boldsymbol{y}$ rather than a multi-dimensional posterior, even if the latter should provide more accurate results. That is to say, we have one KDE for each element in $\boldsymbol{y}$. Moreover, for the second reweighting we use KDEs to update only parameters $\boldsymbol{y}$ but not $\boldsymbol{x}$; this is because, as we shall see below when applying the method on PTA data, the proposal's likelihood cannot be factorized into independent functions of elements in $\boldsymbol{x}$. Hence, one dimensional KDEs for $\boldsymbol{x}$ do not form a good approximation of the target's posterior.


\section{Application in PTA data analyses\label{section 3}}

In this section, we discuss how to apply the likelihood reweighting technique in analyzing the PTA data.  
{We will evaluate the performance of the method by investigating} 
three simulated datasets where a {global sinusoid} signal with various intensities is injected into a SGWB.

\subsection{PTA likelihood}
{Millisecond pulsars are known to have very stable rotations. The times of arrival (TOAs) of the pulses they emit can be recorded fairly precisely by telescopes on earth. The TOAs can then be fit by a timing model, which takes into account the properties of the pulsar and the motion of the solar system and the earth. The differences between the best-fit TOAs and the measured TOAs are known as the residuals.} In the absence of deterministic signals, the residuals are often modeled as the sum of contributions from timing model deviations, (common and/or pulsars' intrinsic) red noise and white noise \cite{vanHaasteren:2008yh,Lentati:2012xb,vanHaasteren:2014qva,taylor2021nanohertz}:
\begin{equation}
\boldsymbol{r}=\boldsymbol{M}\boldsymbol{\epsilon}+\boldsymbol{F}\boldsymbol{a}+\boldsymbol{n}.
\end{equation}
Here $\boldsymbol{r}$ is a concatenated vector composed of the residuals of all pulsars; $\boldsymbol{M}$ is the timing model's design matrix, and $\boldsymbol{\epsilon}$ is a vector of the corresponding coefficients; $\boldsymbol{F}$ represents the Fourier basis of the red noise, and $\boldsymbol{a}$ is a vector of the corresponding amplitudes; $\boldsymbol{n}$ contains white noise from the radiometer, instrumental effects, etc. The PTA likelihood is given by
\begin{equation}
\mathcal{L}(\boldsymbol{r}|\boldsymbol{b})=\frac{\exp\left[-\frac{1}{2}\left(\boldsymbol{r}-\boldsymbol{Tb}\right)^{\top}\boldsymbol{N}^{-1}\left(\boldsymbol{r}-\boldsymbol{Tb}\right)\right]}{\sqrt{\det(2\pi\boldsymbol{N})}},
\end{equation}
where $\boldsymbol{N}$ is the white noise covariance matrix (usually determined by individual single-pulsar analyses), $\boldsymbol{T}=\begin{pmatrix}\boldsymbol{M} & \boldsymbol{F}\end{pmatrix}$ and $\boldsymbol{b}=\begin{pmatrix}\boldsymbol{\epsilon} & \boldsymbol{a}\end{pmatrix}^{\top}$. The prior on $\boldsymbol{b}$ can also be set as Gaussian:
\begin{equation}
\pi(\boldsymbol{b}|\boldsymbol{\eta})=\frac{\exp\left(-\frac{1}{2}\boldsymbol{b}^{\top}\boldsymbol{B}^{-1}\boldsymbol{b}\right)}{\sqrt{\det(2\pi\boldsymbol{B})}}.\label{eq:bB}
\end{equation}
Here the covariance matrix is given by $\boldsymbol{B}=\text{diag}(\infty,\boldsymbol{\phi}(\boldsymbol{\eta}))$, where $\boldsymbol{\eta}$ contains hyperparameters that control $\boldsymbol{B}$. The timing model coefficients $\boldsymbol{\epsilon}$ are well constrained by observations; their inference is likelihood-dominated. Hence we can impose on them a Gaussian prior of infinite variance. As for $\boldsymbol{\phi}(\boldsymbol{\eta})$, in the standard practice, it is assumed to have the form
\begin{equation}
\phi_{({I}i)({{J}}j)}=\left\langle a_{{{I}} {i}}a_{{{J}} j}\right\rangle =\delta_{ij}\left(\delta_{{IJ}}\varphi_{{{I}}i}+\Gamma_{{IJ}}\Phi_{i}\right),
\end{equation}
where $I,J$ range over pulsars and $i,j$ over Fourier components; $\delta_{ij}$ is the Kronecker delta; $\varphi_{{ I}i}$ describes the spectrum of intrinsic red noise in pulsar $I$; and  $\Gamma_{{IJ}}\Phi_{i}$ describes processes with a common spectrum across all pulsars and inter-pulsar correlations. For a SGWB with HD correlations, {which represent the average correlations induced by an isotropic, unpolarized ensemble of gravitational wave sources,} $\Gamma_{ IJ}$ is the HD function of pulsars' angular separations, and $\Phi_{i}$ is usually assumed to obey a power law characterized by amplitude $A_{\rm SGWB}$ and spectral index $\gamma_{\rm SGWB}$:
\begin{equation}
\Phi_{i}=\frac{A_{\rm SGWB}^{2}}{12\pi^{2}}\frac{1}{T}\left(\frac{f_{i}}{1\ \text{yr}^{-1}}\right)^{-\gamma_{\rm SGWB}}1\ \text{yr}^{-3}.\label{eq:power}
\end{equation}
Here $f_{i}$ is the frequency of the $i$th Fourier component and $T$ is the maximum TOAs extent. For a SWGB generated by SMBH binaries, $\gamma$ is expected to be $13/3$ \cite{phinney2001practical}. The intrinsic red noise is usually also assumed to have a power law spectrum like Eq. (\ref{eq:power}); one simply replaces the amplitude $A_{\rm SGWB}$ by $A_I$ and the spectral index $\gamma_{\rm SGWB}$ by $\gamma_I$ for each pulsar. 

The full hierarchical PTA posterior can then be written as
\begin{equation}
p(\boldsymbol{b},\boldsymbol{\eta}|\boldsymbol{r})=\mathcal{L}(\boldsymbol{r}|\boldsymbol{b})\pi(\boldsymbol{b}|\boldsymbol{\eta})\pi(\boldsymbol{\eta}),
\end{equation}
where $\pi(\boldsymbol{\eta})$ is the hyperprior on $\boldsymbol{\eta}$. Compared with hyperparameters $\boldsymbol{\eta}$, e.g., the amplitude ($A_{\rm SGWB}$) and the spectral index ($\gamma_{\rm SGWB}$) of the SGWB power spectrum, parameters $\boldsymbol{b}$, i.e.,  the coefficients of the design matrix $\boldsymbol{\epsilon}$ and the Fourier coefficients $\boldsymbol{a}$, are generally not of particular interest. Note that the hierarchical likelihood $\mathcal{L}(\boldsymbol{r}|\boldsymbol{b})\pi(\boldsymbol{b}|\boldsymbol{\eta})$ is a Gaussian function for $\boldsymbol{b}$. One can then integrate out these parameters analytically and obtain the marginalized likelihood that only depends on $\boldsymbol{\eta}$ \cite{Lentati:2012xb,vanHaasteren:2012hj}:
\begin{align}
\mathcal{L}_{\rm SGWB}(\boldsymbol{r}|\boldsymbol{\eta})  & =\int\mathcal{L}(\boldsymbol{r}|\boldsymbol{b})\pi(\boldsymbol{b}|\boldsymbol{\eta})\text{d}\boldsymbol{\epsilon}\text{d}\boldsymbol{a}\\ & =\frac{\exp\left(-\frac{1}{2}\boldsymbol{r}^{\top}\boldsymbol{C}^{-1}\boldsymbol{r}\right)}{\sqrt{\det(2\pi\boldsymbol{C})}},\label{eq:marginalized likelihood}
\end{align}
where $\boldsymbol{C}=\boldsymbol{N}+\boldsymbol{T}\boldsymbol{B}\boldsymbol{T}^{\top}$, and we have used the Woodbury identity that gives $\boldsymbol{C}^{-1}=\boldsymbol{N}^{-1}+\boldsymbol{N}^{-1}\boldsymbol{T}\left(\boldsymbol{B}^{-1}+\boldsymbol{T}^{\top}\boldsymbol{N}^{-1}\boldsymbol{T}\right)^{-1}\boldsymbol{T}^{\top}\boldsymbol{N}^{-1}$. This is practically the likelihood implemented within production-level GW search pipelines, such as $\tt enterprise$ \cite{enterprise}.\footnote{\url{https://github.com/nanograv/enterprise}} The bottleneck in computing this likelihood comes from the inversion of the matrix $\boldsymbol{C}$, whose dimension is the total number of TOAs in the PTA.

{We may consider two types of possible deviations from the fiducial HD pattern: (1) Global effects, such as clock errors (which generate a monopole correlation) and gravity theories beyond general relativity, which result in different inter-pulsar correlations, and thus bring modifications to $\boldsymbol{\phi}$ (in the covariance matrix $\boldsymbol{C}$). (2) Deterministic signals that stand out of the SGWB, such as strong continuous waves from two inspiraling SMBHs, which alter the residuals $\boldsymbol{r}$ by subtracting the signal-induced residuals: $\boldsymbol{r}\to\boldsymbol{r}- \boldsymbol{r}_{\text{det}}$. If these deviations are small, parameters describing the new features are not expected to significantly impact the dataset compared to, e.g., the common red noise parameters. In this case, a model containing an HD SGWB can be regarded as a proposal model for likelihood reweighting, based on which we study any target models we are interested in.}

\subsection{SGWB + global sinusoid signal}

Let us consider a deterministic sinusoid signal applied to all pulsars. Such a signal can be the result of perturbations to the local gravitational field induced by ultralight scalar dark matter \cite{Khmelnitsky:2013lxt}. It was specifically investigated in NANOGrav's 15-year analysis, because a slight enhancement in the SGWB power spectrum was found at frequency $\sim 3.95$ nHz from a frequentist analysis (although no significant evidence was found in the Bayesian approach) \cite{NANOGrav:2023gor}. Here we are using the signal as an exemplar to demonstrate likelihood reweighting.

The datasets we explored contain 20 pulsars randomly distributed on the sky map with an observation span of 10 years.\footnote{The datasets were generated by $\tt libstempo$  (\url{github.com/vallis/libstempo}), a Python wrapper around the $\tt tempo2$ pulsar timing package.} TOAs were taken once per month. Each pulsar is subject to white noise and power law intrinsic red noise. We fixed the white noise level to be $1\ \mu\text{s}$. As for the intrinsic red noise, the amplitude and the spectral index for each pulsar have injected values randomly drawn from their priors listed in Table \ref{tab:priors}.  In addition, we injected a SGWB of power law spectrum with $A_{\rm SGWB}=10^{-14}$  s and $\gamma_{\rm SGWB}=13/3$. Although the SGWB induces HD correlations, we treated the background as common uncorrelated red noise (CURN). This is known to significantly reduce the computational cost because the noise matrix $\boldsymbol{C}$ is now block-diagonal, which allows $\boldsymbol{C}^{-1}$ to be computed block by block {(or pulsar by pulsar)}. In addition to the noise, we injected a sinusoidal signal into all pulsars. The induced residuals are determined by three parameters: amplitude $A_{\rm sin}$, frequency $f_{\rm sin}$, and phase $\phi_{\rm sin}$. For a certain TOA $t$, the induced residual is
\begin{equation}
r_{\rm sin}=A_{\rm sin}\sin\left(2\pi f_{\rm sin}t+\phi_{\rm sin}\right).
\end{equation}
In total, there are $20\times 2 + 2 + 3= 45$ parameters, whose injected values are listed in Table \ref{tab:priors}. 

We generated three datasets and the only difference was the value of $A_{\rm sin}$. For each dataset, we performed two full Bayesian analyses: $\textsf{CURN}$ and $\textsf{CURN+sin}$; the former does not include the sinusoid signal. The reweighting was tested based on $\textsf{CURN}$ (42 parameters), which was considered as the proposal model. The reweighted samples were compared with the $\textsf{CURN+sin}$ results. The likelihoods of the two models are
\begin{equation}
\mathcal{L}_{\text{CURN}}=\frac{\exp\left(-\frac{1}{2}\boldsymbol{r}^{\top}\boldsymbol{C}^{-1}\boldsymbol{r}\right)}{\sqrt{\det(2\pi\boldsymbol{C})}},
\end{equation}
\begin{equation}
\mathcal{L}_{\text{CURN+sin}}=\frac{\exp\left[-\frac{1}{2}\left(\boldsymbol{r}-\boldsymbol{r}_{\rm sin}\right)^{\top}\boldsymbol{C}^{-1}\left(\boldsymbol{r}-\boldsymbol{r}_{\rm sin}\right)\right]}{\sqrt{\det(2\pi\boldsymbol{C})}}.
\end{equation}
The priors of all parameters are listed in Table \ref{tab:priors}. In this work, these likelihoods were computed by $\tt enterprise$, and the MCMC Bayesian analyses were carried out by $\tt enterprise\_extension$\footnote{\url{https://github.com/nanograv/enterprise_extensions}} \cite{enterprise_extension} and $\tt PTMCMCSampler$\footnote{\url{https://github.com/nanograv/PTMCMCSampler}} \cite{justin_ellis_2017_1037579}. To monitor chain convergence, we adopted the concept of Gelman-Rubin statistic \cite{brooks1998general}. Since the PTA likelihood is usually expensive to compute, only one MCMC chain was obtained for each analysis, rather than multiple; then we divided the chain in half after the burn-in {(the first quarter of the samples in our case)} and used the two sub-chains to compute the Gelman-Rubin statistic. We say a chain is converged when the $R$ value is smaller than 1.005 for all parameters.

{The values of $\log_{10}A_{\rm sin}$ in the three datasets are $-6.8,-6.7$ and $-6.6$, respectively. Despite the small differences in the signal amplitude, as we will see, the corresponding} Bayes factors (favoring $\textsf{CURN+sin}$ over $\textsf{CURN}$) are $\mathcal{B}\sim 1, \sim 10$ and $\sim 500$. For convenience we shall refer to the signals as weak, moderate and strong.

\begin{table*}[ht]
\begin{center}
\scriptsize
\caption{Prior distributions and injected values used in analyses performed in this paper.}
\label{tab:priors}
\begin{tabular}{llll}
\hline\hline
{parameter} & {description} & {prior} & {injected value} \\
\hline

\multicolumn{4}{c}{\textit{power-law intrinsic red noise of the $I$th pulsar}} \\[1pt]
$A_{I}$ & amplitude& log-uniform $[-18, -11]$ & drawn from prior  \\
$\gamma_{I}$ & spectral index & Uniform $[0, 7]$ & drawn from prior \\
\hline

\multicolumn{4}{c}{\textit{power-law CURN}} \\[1pt]
$A_{\rm SGWB}$ & amplitude & log-uniform $[-18, -11]$ & $10^{-14}$ \\
$\gamma_{\rm SGWB}$ & spectral index & Uniform $[0,7]$ & $13/3$ \\
\hline

\multicolumn{4}{c}{\textit{sinusoid signal}} \\[1pt]
$A_{\rm sin}$ [s] & amplitude & log-uniform $[-10, -5]$ & weak: $10^{-6.8}$\\ 
& & & moderate: $10^{-6.7}$\\
& & & strong$10^{-6.6}$ \\
$f_{\rm sin}$ [Hz] & frequency & log-uniform $[-8.5,-7.5]$ & $10^{-8}$ \\
$\phi_{\rm sin}$ & phase & Uniform $[0,2\pi]$ & $\pi$ \\
\hline

\end{tabular}
\end{center}
\end{table*}

\subsubsection{Weak signal}

\begin{figure}
\centering
\vspace{0.1in}
\includegraphics[scale=0.5]{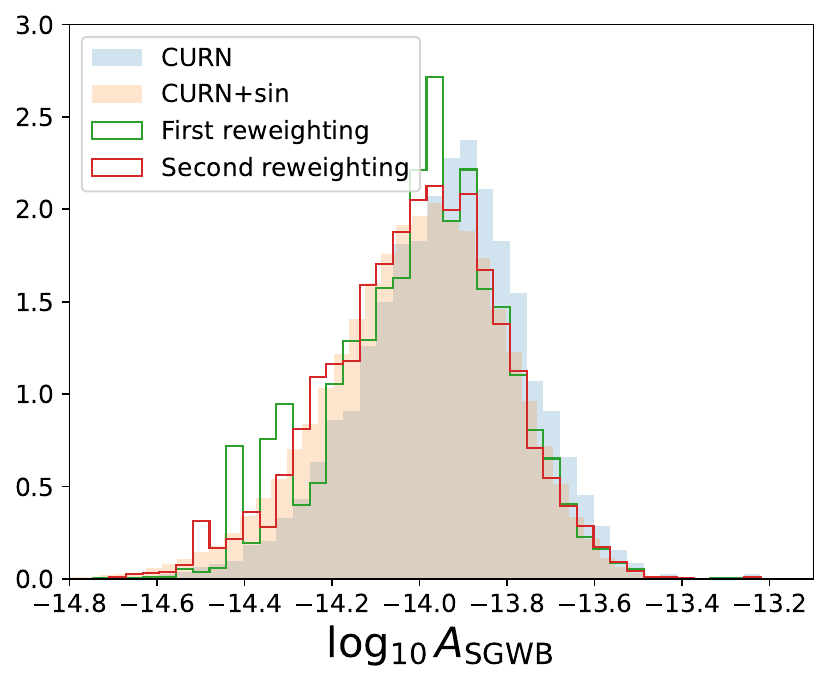}
\includegraphics[scale=0.5]{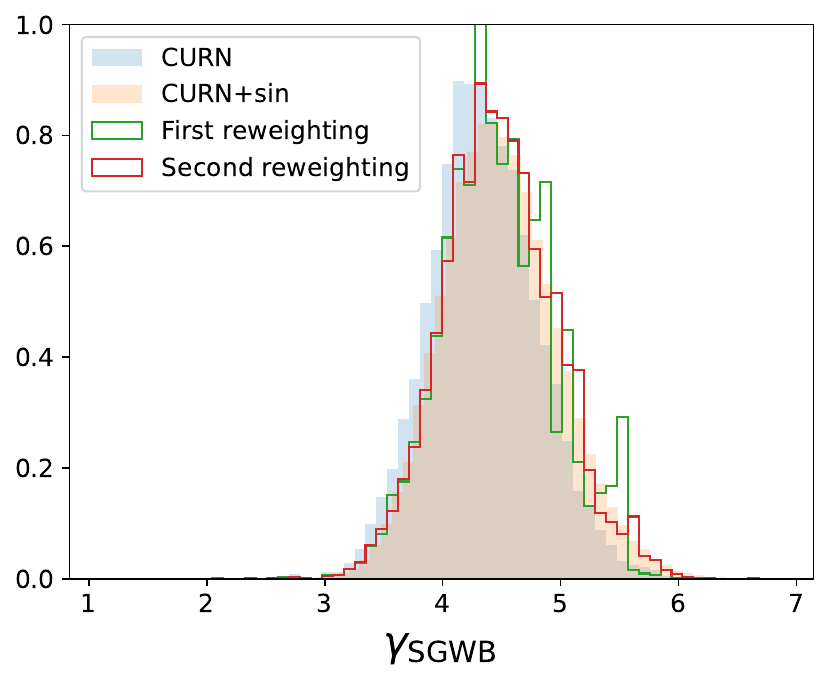}

\caption{\label{fig:weak1} ({\it Weak signal scenario}) Marginal posteriors of the SGWB parameters $A_{\rm SGWB}$ and $\gamma_{\rm SGWB}$ from full Bayesian runs of $\mathsf{CURN}$ and $\mathsf{CURN+sin}$, as well as the first and the second reweightings. The posteriors are faithful reconstructed by the reweighting technique. The injected values are $\log_{10}A_{\rm SGWB}=-14$ and $\gamma_{\rm SGWB}=13/3\approx 4.33$.}
\end{figure}

We first considered a dataset where the sinusoid signal is almost buried by noise. Fig. \ref{fig:weak1} shows the marginal posteriors of $\log_{10} A_{\rm SGWB}$ and $\gamma_{\rm SGWB}$ in $\mathsf{CURN}$ and $\mathsf{CURN+sin}$. We do not see much difference between the two models, which means that the presence of the sinusoid signal does not bias the search for the SGWB.

\begin{figure}
\centering
\vspace{0.1in}
\includegraphics[scale=0.5]{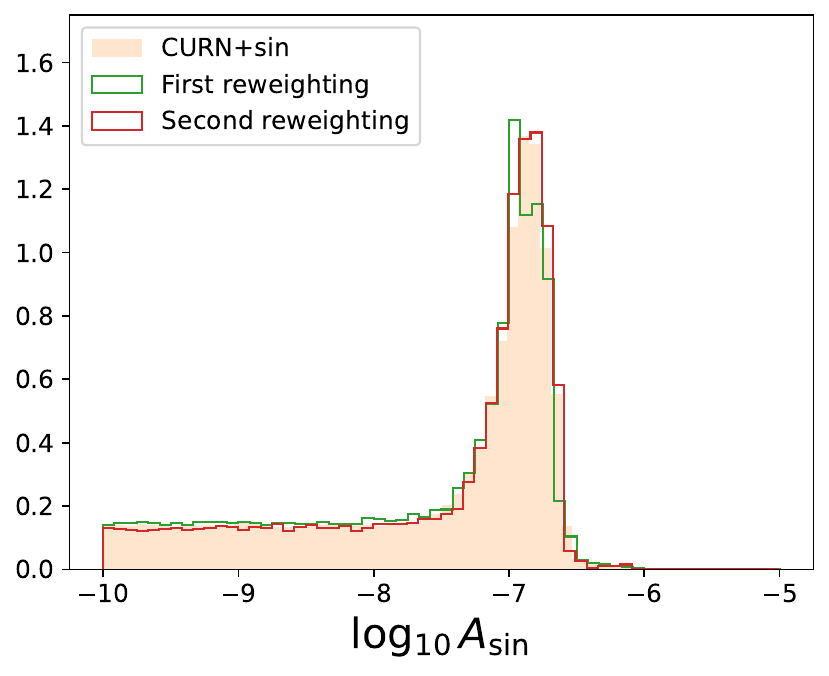}
\includegraphics[scale=0.5]{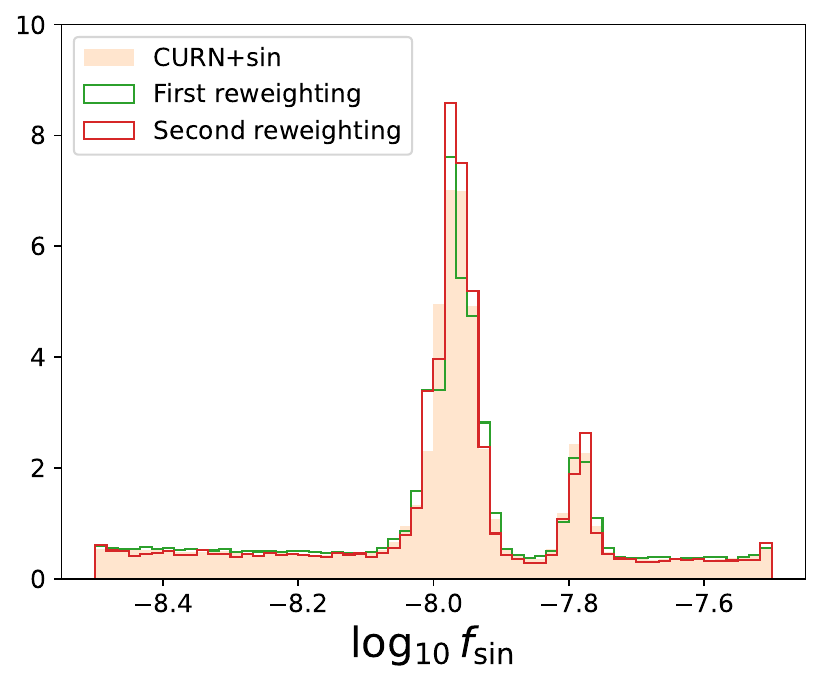}

\caption{\label{fig:weak2} ({\it Weak signal scenario}) Marginal posteriors of the signal parameters $A_{\rm sin}$ and $f_{\rm sin}$ from a full Bayesian run, as well as the first and the second reweightings based on $\mathsf{CURN}$. For such a weak signal, the posteriors are faithful reconstructed by the reweighting technique. The injected values are $\log_{10}A_{\rm sin}=-6.8$ and $\log_{10}f_{\rm sin}=-8$.}
\end{figure}

It is also shown in Fig. \ref{fig:weak1} the reweighted posteriors of $\log_{10}A_{\rm SGWB}$ and $\gamma_{\rm SGWB}$ after reweighting once. Unsurprisingly, up to some unstable weights, they agree with the results from an MCMC search. Fig. \ref{fig:weak2} shows the posteriors of $\log_{10}A_{\rm sin}$ and $\log_{10}f_{\rm sin}$ from $\mathsf{CURN+sin}$ and from reweighting. Both sets of posteriors almost overlap.

We then carried out the reweighting for the second time using the KDE technique described in the previous section.\footnote{We used $\tt scipy$ (scipy.stats.gaussian\_kde) \cite{virtanen2020scipy} for Eq. \ref{fKED} and for sampling from $p_{\rm KDE}(\boldsymbol{y})$, and then used $\tt KDEpy$ \cite{odland2018tommyod} to evaluate $f_{\rm KDE}$ of these samples in the weight calculation.} Specifically, we took the KDEs of the sinusoid parameters' one dimensional marginal posteriors [Eq. (\ref{fKED})], which were then treated as three independent likelihoods $f_{\rm KDE}^{(1)}(A_{\rm sin})$, $f_{\rm KDE}^{(2)}(f_{\rm sin})$ and $f_{\rm KDE}^{(3)}(\phi_{\rm sin})$ . Along with the $\mathsf{CURN}$ likelihood, they formed the second proposal model with likelihood $\mathcal{L}_{\rm CURN}f_{\rm KDE}^{(1)}f_{\rm KDE}^{(2)}f_{\rm KDE}^{(3)}$. Using the KDEs, we drew posterior samples assuming uniform priors as in $\mathsf{CURN+sin}$. These samples were used in Eqs. (\ref{eq:pos_x_y2}) and (\ref{Z2}) in obtaining the weights and reweigted samples. The reweighted posteriors are shown in Figs. \ref{fig:weak1} and \ref{fig:weak2}.  We can see that the reweighted parameters are now more compatible with the results found in a $\mathsf{CURN+sin}$ search.

For model selection, we computed the Bayes factor using the hypermodel method \cite{hee2016bayesian}; an MCMC sampler using this method was implemented in $\tt enterprise\_extensions$.\footnote{One can also use the marginal posterior of $\log_{10} A_{\rm sin}$ to estimate the Savage-Dickey Bayes factor \cite{dickey1971weighted} because $\mathsf{CURN+sin}$ reduces to $\mathsf{CURN}$ when $A_{\rm sin}$ takes a small value.} The result is $\mathcal{B}=1.48\pm 0.04$. On the other hand, the second reweighting gives $\mathcal{B}=1.63\pm 0.03$. Although the posteriors of $\log_{10} A_{\rm sin}$ and $\log_{10} f_{\rm sin}$ do show some inclination towards the injected values, in reality we cannot call detection due to the small Bayes factor.\footnote{NANOGrav found $\mathcal{B}\approx 1$ in the 15-year data analysis supporting $\mathsf{HD+sin}$ over $\mathsf{HD}$ \cite{NANOGrav:2023gor}.}

We see that implementing likelihood reweighting gives us almost the same results as those from a full Bayesian run. The method serves as a good tool for a screening test on a specific model beyond the $\mathsf{CURN}$ model. For a weak signal, implementing reweighting once is sufficient to extract the information about the model parameters (so that one can impose upper limits on the signal). If the Bayes factor (the mean of weights) is too small, the model is not very well constrained by the dataset.

\subsubsection{Moderate signal}

\begin{figure}
\centering
\vspace{0.1in}
\includegraphics[scale=0.5]{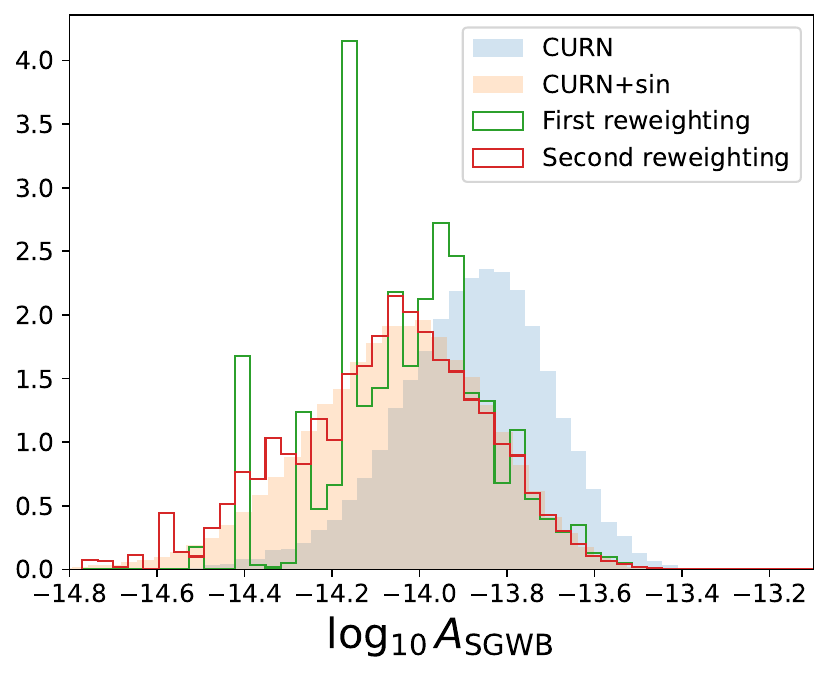}
\includegraphics[scale=0.5]{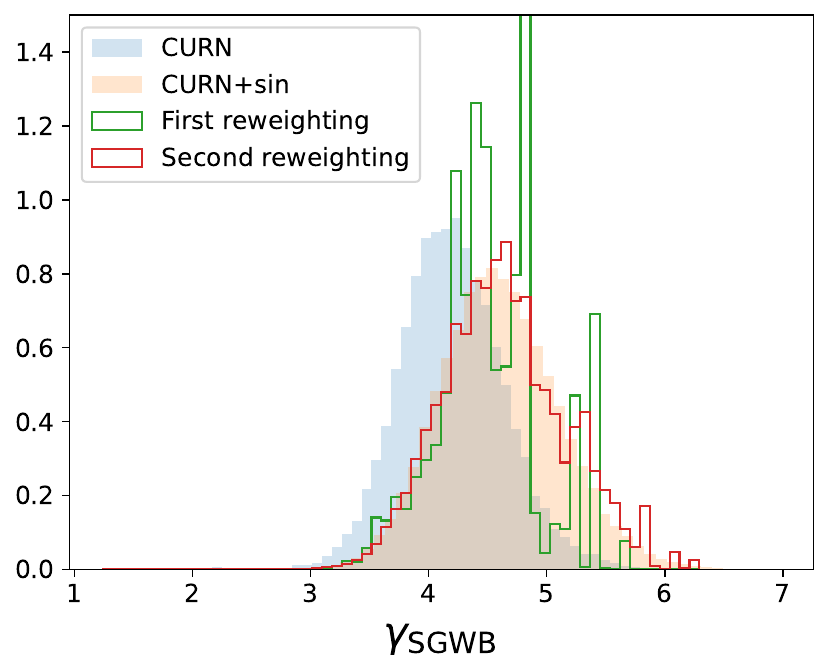}

\caption{\label{fig:moderate1} ({\it Moderate signal scenario}) Marginal posteriors of the SGWB parameters $A_{\rm SGWB}$ and $\gamma_{\rm SGWB}$ from full Bayesian runs of $\mathsf{CURN}$ and $\mathsf{CURN+sin}$, as well as the first and the second reweightings. While the first reweighting gives unstable weight estimates, the results of the second reweighting are compatible with the Bayesian search. The injected values are $\log_{10}A_{\rm SGWB}=-14$ and $\gamma_{\rm SGWB}=13/3\approx 4.33$.}
\end{figure}

Now we consider a stronger signal. The two models are slightly disjoint in this regime. Fig. \ref{fig:moderate1} shows that the existence of the sinusoid signal obviously biases the estimate of the $\mathsf{CURN}$ parameters. In particular, $A_{\rm SGWB}$ in the $\mathsf{CURN}$ model is overestimated because the sinusoid was mistaken as part of the background. It is also shown in Fig. \ref{fig:moderate1} the reweighted posteriors after reweighting once. As expected, larger weights are assigned to samples closer to the injected values. However, there are not sufficient samples at the tails of the $\mathsf{CURN}$ posteriors, which causes unstable weight estimates.

Following the same procedure as described in the previous subsection, we then performed the reweighting for the second time. The reweighted posteriors are also shown in Fig. \ref{fig:moderate1}. We can see that the reweighted SGWB parameters are now more compatible with the results found in a $\mathsf{CURN+sin}$ search.

\begin{figure}
\centering
\vspace{0.1in}
\includegraphics[scale=0.5]{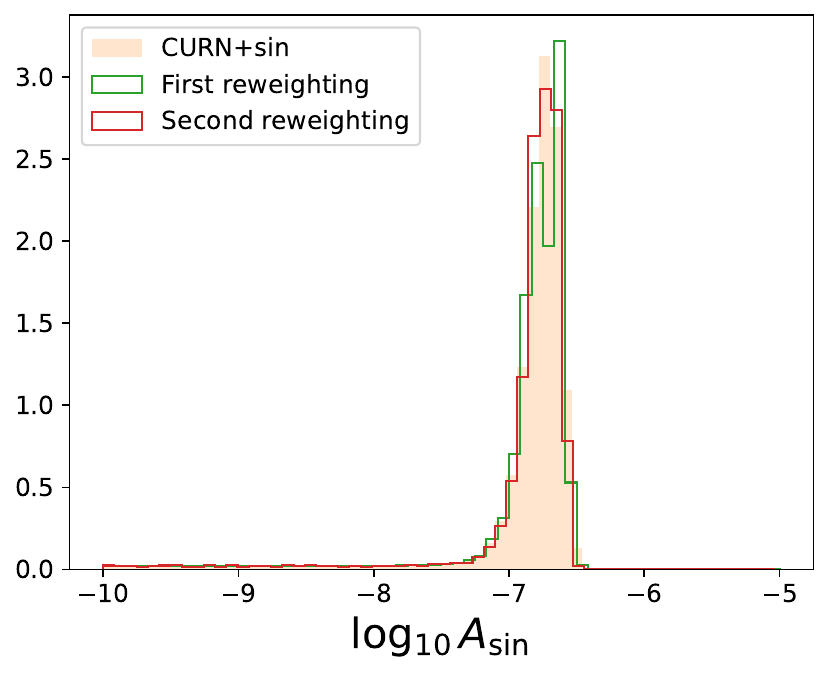}
\includegraphics[scale=0.5]{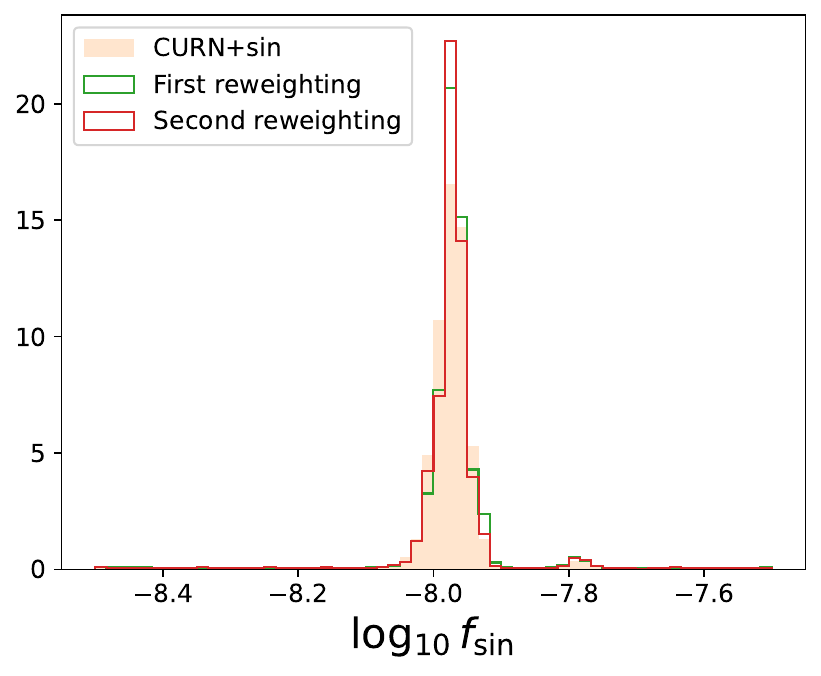}

\caption{\label{fig:moderate2}({\it Moderate signal scenario}) Marginal posteriors of the signal parameters $A_{\rm sin}$ and $f_{\rm sin}$ from a full Bayesian run, as well as the first and the second reweightings based on $\mathsf{CURN}$. For this moderate signal (the Bayes factor between $\mathsf{CURN+sin}$ and $\mathsf{CURN}$ is $\sim 10$), the first reweighting is able to faithfully recover the signal posteriors. The injected values are $\log_{10}A_{\rm sin}=-6.7$ and $\log_{10}f_{\rm sin}=-8$.}
\end{figure}

In Fig. \ref{fig:moderate2} we show the signal posteriors from a $\mathsf{CURN+sin}$ search, the first reweighting and the second reweighting. Interestingly, unlike the case for the SGWB parameters, the first reweighting already resulted in decent reconstruction of the injected signal. The reason behind this could be that the deterministic signal is now so strong that the poor reconstruction of the SGWB parameters does not bias the search much.

A hypermodel run of this dataset gives a Bayes factor $\mathcal{B}=10.16\pm 0.74$, while the second reweighting gives $\mathcal{B}=11.07\pm 0.23$. This indicates strong (but not decisive) evidence for the signal over pure $\mathsf{CURN}$.

\subsubsection{Strong signal}

As expected, likelihood reweighting serves as an efficient tool in investigating a target model that can be considered as a ``perturbed" proposal model. Now we turn to the case where the signal dominates over the common process. The amplitude of the sinusoid is sufficiently large so that the Bayes factor is $\mathcal{B}\approx 450$, which can be regarded as decisive evidence in favor of the existence of the signal beyond CURN.

\begin{figure}
\centering
\vspace{0.1in}
\includegraphics[scale=0.5]{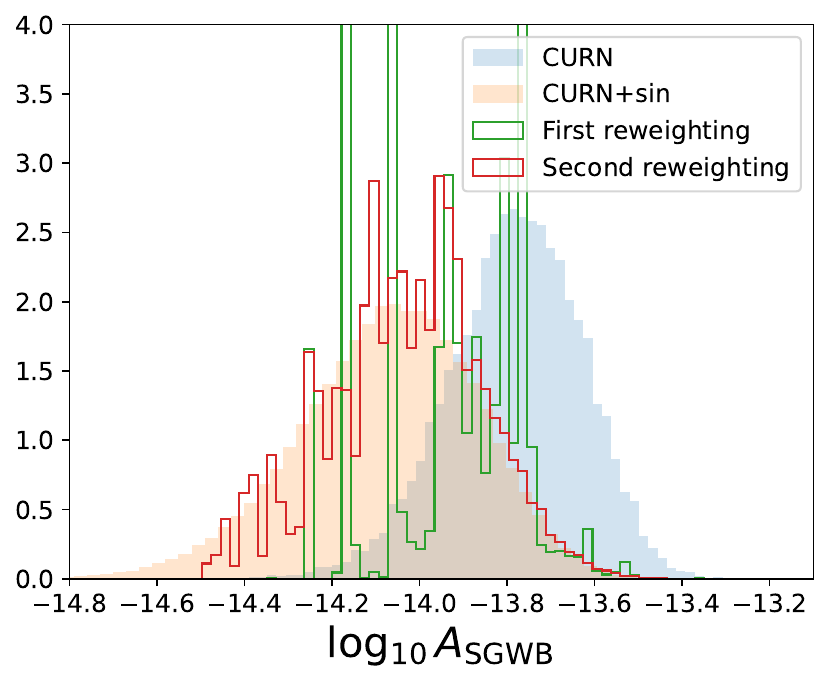}
\includegraphics[scale=0.5]{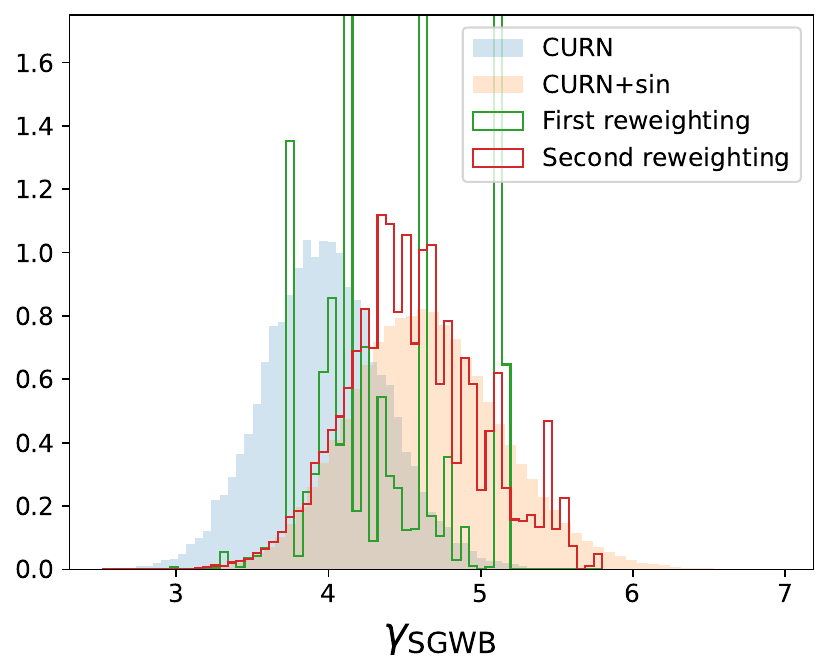}

\caption{\label{fig:strong1} ({\it Strong signal scenario}) Marginal posteriors of the SGWB parameters $A_{\rm SGWB}$ and $\gamma_{\rm SGWB}$ from full Bayesian runs of $\mathsf{CURN}$ and $\mathsf{CURN+sin}$, as well as the first and the second reweightings. We can still see features of unstable weights after the second reweighting. The injected values are $\log_{10}A_{\rm SGWB}=-14$ and $\gamma_{\rm SGWB}=13/3\approx 4.33$.}
\end{figure}

\begin{figure}
\centering
\vspace{0.1in}
\includegraphics[scale=0.5]{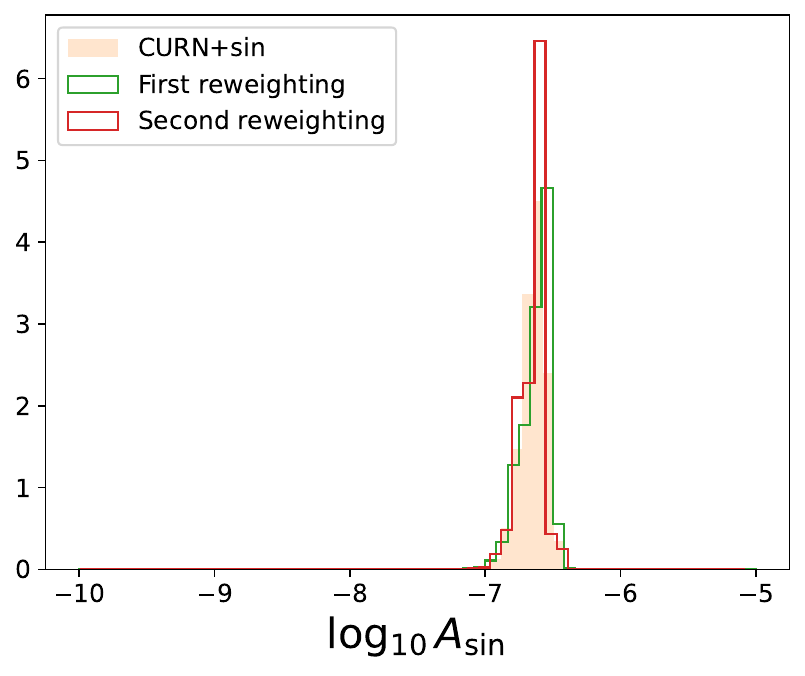}
\includegraphics[scale=0.5]{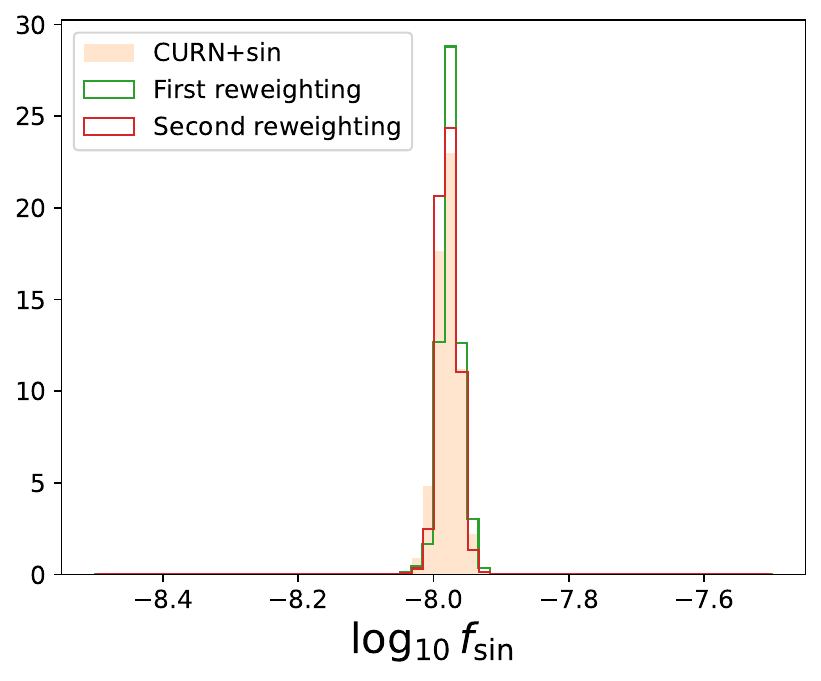}

\caption{\label{fig:strong2} ({\it Strong signal scenario}) Marginal posteriors of the signal parameters $A_{\rm sin}$ and $f_{\rm sin}$ from a full Bayesian run, as well as the first and the second reweightings based on $\mathsf{CURN}$. The injected signal can be captured by the first reweighting. The injected values are $\log_{10}A_{\rm sin}=-6.6$ and $\log_{10}f_{\rm sin}=-8$.}
\end{figure}

As we can see from Fig. \ref{fig:strong1}, even after reweighting twice, features of ``fluctuating" weights persist in the SGWB posteriors.\footnote{Of course, the problem can be alleviated if we choose more samples from $\mathsf{CURN}$ for reweighting, which does not require much more effort in our tests. Here we are showing how the effect of unstable weights manifests with the least possible number of samples from the proposal's chain.} However, if we look at the signal posteriors in Fig. \ref{fig:strong2}, even after the first reweighting, when we have a poor reconstruction of the SGWB posteriors, the injected signal is well captured. The signal is now so dominant that the SGWB parameters become more or less irrelevant. 

Samples at small $A_{\rm sin}$ having tiny weights implies decisive evidence for the presence of the signal. In fact, a hypermodel run gives a Bayes factor $\mathcal{B}=448\pm 42$, and the second reweighting gives $\mathcal{B}=491\pm 23$. This immediately demands a full analysis of the target model. Reweighting applied on such a dataset can quickly tell us that the fiducial model is strongly disfavored, and that one should turn to something beyond.\newline 

{When performing an MCMC $\mathsf{CURN+sin}$ search, most runtime is ``wasted" in searching for the ``$\mathsf{CURN}$" part rather than the ``$\mathsf{sin}$" part. Now let us estimate the speedup that likelihood reweighting brings in general. For simplicity, we assume that the sample number needed for a converged MCMC chain in the target is equal to that in the proposal; we also assume that the thin-out of the proposal's chain is carried out every 10 samples.\footnote{This is the default in \tt{enterprise}.} In the case where the target's likelihood is much more expensive than the proposal, as in, e.g., Refs. \cite{Payne:2019wmy,Romero-Shaw:2019itr}, one obviously gets a $\mathcal{O}(10)$-time speedup. If the target's likelihood is not that difficult to compute, implementing likelihood reweighting may seem unnecessary, since we also need to account for the time it takes to obtain the proposal's chain. However, when facing a PTA dataset, one always performs a search for the HD SGWB anyways. The resulting samples can help us screen a number of targets. Furthermore, computing the weights in parallel can reduce the expense significantly. In this sense, we conclude that likelihood reweighting could bring an at least $\mathcal{O}(10)$-time speedup in dealing with PTA datasets.}

\section{Conclusions and discussion\label{section 4}}

In this paper, we have further investigated the method of likelihood reweighting applied in PTA datasets. Following PTA collaborations' reported evidence supporting the existence of a SGWB in the Universe, a natural question is: Are there features beyond the fiducial HD model that we may see in PTA datasets? If these new features exist in nature, one may expect them to behave like ``perturbations" to the fiducial model in the upcoming datasets in the near future. {In this case,} given posterior samples from the fiducial model, likelihood reweighting provides an at least $\mathcal{O}(10)$-time speedup in exploring more sophisticated models compared to full Bayesian searches. 

This method is well known to suffer from sampling error if the fiducial model and the target model are disjoint. Inspired by adaptive importance sampling, we proposed reweighting for the second time based on the results from the first reweighting. A likelihood obtained by KDE is assigned to the new parameters, which, along with the likelihood in the fiducial model, form a new proposal model. The second reweighting is then based on this new likelihood. This is expected to alleviate the vanilla importance sampling's error because the second proposal already has preference for the new parameters.

We tested the method using three datasets with a global sinusoid signal injected into the SGWB. We found that likelihood reweighting not only behaves decently when the signal is weak, but is also able to capture a strong signal even if the SGWB parameters' posterior is poorly recovered. In the weak-signal regime, the fiducial model is a good approximation of the target, when likelihood reweighting is expected to work well; in the strong-signal regime, the method fails with unstable weights, but the signal can more or less be extracted because the target likelihood is signal-dominated. Furthermore, we found that the Bayes factors obtained by the averaged weights are consistent with those from a formal method (hypermodel search in our {experiments}). 

Note that the performance of likelihood reweighting can be further improved in our experiments if we assign more than one weight to each proposal's sample. This can be easily done by drawing more than one random sample of $\boldsymbol{y}$ for each $\boldsymbol{x}^{(i)}$ in the first reweighting. The trade-off is the increase in runtime. However, this step is applicable if we expect that the convergence of the target's chain requires a large number of MCMC samples (larger than the proposal's chain).

{In general, likelihood reweighting can serve as an efficient tool to screen new models. We can consider two extreme scenarios where the technique is particularly useful: (1) Both the Bayes factor and its error are small. This means that the signal is weak, and that constraints (upper limits) can be imposed on the model. (2) Both the Bayes factor and its error are relatively large. This immediately encourages us to further investigate the new model with a more robust search.}

We stress that, for the first reweighting, samples of the new parameters are usually drawn from a uniform distribution, which is basically vanilla Monte Carlo. If there are too many new parameters, random sampling cannot catch all the important features of the likelihood. Therefore, we do not expect the technique to work properly if the number of new parameters is relatively large ($\gtrsim 10$). A simple and physically interesting extension of the current work is to consider a sinusoid signal emitted from a source on the sky map. The impact of the signal on the residuals depends on each pulsar's sky location. This model is akin to continuous waves from a SMBH binary (with only the ``earth term" taken into consideration). It has two more parameters (characterizing the sky location of the source) in addition to the sinusoid we studied in this work.

Lastly, although in this work we performed likelihood reweighting based on a SGWB model ($\mathsf{CURN}$), it can be envisioned that a more sophisticated model can be adopted as the proposal. For example, in NANOGrav's 15-year analysis, the $\mathsf{HD+sin}$ model was explored through likelihood reweighting based on $\mathsf{CURN+sin}$, where the latter is much easier to sample from. The parameter numbers in the two models are the same, and reweighting for the second time is not necessary. Another possible application of the technique is to look for the eccentricity of a SMBH binary. Accounting for the eccentricity in the search for continuous waves is known to lead to an expensive likelihood. Adopting the reweighting method based on results from a circular-orbit search may bring valuable insights at a much lower computational cost. This will be further studied in a future work \cite{Eccentricity:2024}.    

\section*{Acknowledgments}
We are grateful to Ken Olum for bringing to our attention the stepping-stone algorithm. We also thank Amanda Malnati, Patrick Meyers and Chiara Mingarelli for insightful comments on the manuscript. This project received support from the National Science Foundation (NSF) Physics Frontiers Center award numbers 1430284 and 2020265. HD was also supported by Yuri Levin’s Simons Investigator Grant PG012519.

\bibliography{likelihood_reweighting_sinusoid}

\end{document}